\documentclass[conference]{IEEEtran}
\IEEEoverridecommandlockouts

\usepackage{amsthm}
\newtheorem{theorem}{Theorem}

\newtheorem{example}{Example}
\newtheorem*{remark}{Remark}
\usepackage{cite}
\usepackage{amsmath,amssymb,amsfonts}
\usepackage{algorithmic}
\usepackage{graphicx}
\usepackage[english]{babel}
\usepackage{textcomp}
\usepackage{xcolor}

\ifCLASSOPTIONcompsoc
    \usepackage[caption=false, font=normalsize, labelfont=sf, textfont=sf]{subfig}
\else
\usepackage[caption=false, font=footnotesize]{subfig}
\fi

\begin{document}

\title{Achieving Short-Blocklength RCU bound
via CRC List Decoding of TCM with Probabilistic Shaping\\
}

\author{\IEEEauthorblockN{Linfang Wang\IEEEauthorrefmark{1},
Dan Song\IEEEauthorrefmark{1},
Felipe Areces\IEEEauthorrefmark{1},
Richard D. Wesel\IEEEauthorrefmark{1}}
\IEEEauthorblockA{\IEEEauthorrefmark{1}University of California, Los Angeles, Los Angeles, CA 90095, USA}
Email: \{lfwang,dansong,fareces99,wesel\}@ucla.edu  
}

\maketitle

\begin{abstract}
This paper applies probabilistic amplitude shaping (PAS) to a cyclic redundancy check (CRC) aided trellis coded modulation (TCM) to achieve the short-blocklength random coding union (RCU) bound. 
In the transmitter, the equally likely message bits are first encoded by distribution matcher to generate amplitude symbols with the desired distribution. 
The binary representations of the distribution matcher outputs are then encoded by a CRC. 
Finally, the CRC-encoded bits are encoded and modulated by Ungerboeck's TCM scheme, which consists of a systematic $\frac{k_0}{k_0+1}$ tail-biting convolutional code and a mapping function that maps coded bits to channel signals with capacity-achieving distribution. 
This paper proves that, for the proposed transmitter, the CRC bits have uniform distribution and that the channel signals have symmetric distribution.
In the receiver, the serial list Viterbi decoding (S-LVD) is used to estimate the information bits. 
Simulation results show that, for the proposed CRC-TCM-PAS system with 87 input bits and 65-67 8-AM coded output symbols, the decoding performance under additive white Gaussian noise channel achieves the RCU bound with properly designed CRC and convolutional codes. 
\end{abstract}


{\let\thefootnote\relax\footnote{{This research is supported by National Science Foundation (NSF) grants CCF-1911166 and CCF-2008918. Any opinions, findings, and conclusions or recommendations expressed in this material are those of the author(s) and do not necessarily reflect views of NSF.}}}
\section{Introduction}
On the additive white Gaussian noise (AWGN) channel, the spectral efficiency can be improved by the probabilistic shaping (PS)\cite{pas_bocherer_1} which optimizes the shape and probability mass function (PMF) of the constellation set. The combination of PS and forward error correction (FEC) techniques further boosts the performance of a transmission system\cite{pas_bocherer_1,pas_bocherer_2,pas_achievablerate_bocherer,polar_ook_thomas}. A well-known layered PS architecture is probabilistic amplitude shaping (PAS) \cite{pas_bocherer_1,pas_bocherer_2}.

The transmitter of a PAS consists three modules. The first module is a distribution matcher (DM) which maps a sequence of binary bits with uniform distribution to a sequence of magnitude symbols that obey a desired distribution. In practice, DM cannot generate symbols that have arbitrary distribution because of finite input length. Hence, a good DM should generate symbols with distribution that is as close as to the desired one. 
The popular DMs are  shell-mapping (SM) DM\cite{SMDM_amjad,ccdm_smdm_comp_schulte}, constant composition (CC) DM \cite{CCDM_schulte} and other forms of DM \cite{DMReview}.



The second module of PAS architecture is FEC. The parity check bits of FEC serve as the sign sequence for the magnitude sequence. The last module realizes modulation by entrywise multiplication between the sign sequence and the magnitude sequence. The authors in \cite{pas_bocherer_1,CCDM_schulte} use low-density parity-check (LDPC) codes as error correction codes. 


\begin{figure*}[t] 
\centering
  \includegraphics[scale = 0.52]{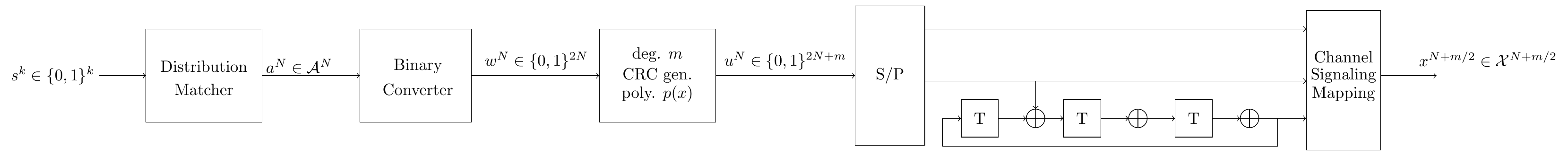}
  \caption{Diagram of CRC-TCM-PAS transmitter with 8-AM coded modulation.}
  \label{fig: transmitter}
\end{figure*}

A FEC code with excellent error correction performance is vital for PAS transmission system. LDPC code has shown near-capacity performance with long blocklength\cite{LDPC_capacity}. However LDPC codes with short blocklength do not perform as well as LDPC codes with long blocklength, as a result of short cycles in the Tanner graph corresponding to the code.

Recently, convolutional codes concatenated with CRCs have shown excellent performance in short blocklength regimes\cite{Liang_tbcccrc,yang2018serial,yang2019list}. Yang \textit{et al.} in \cite{Liang_tbcccrc} show that a tail-biting convolutional code (TBCC) concatenated with CRC can achieve FER performance very close to random short-blocklength RCU bound when the decoder implements the serial list Viterbi decoding (S-LVD) algorithm.   

This paper CRC-TCM-PAS architecture, which applies probabilistic amplitude shaping  to a cyclic redundancy check  aided trellis coded modulation to achieve the short-blocklength RCU bound. 
In contrast with previous works \cite{pas_bocherer_1,pas_bocherer_2,ccdm_smdm_comp_schulte}, which use LDPC codes as FEC to provide sign sequences for the input magnitude sequence, this paper proposes a CRC-aided trellis coded modulation\cite{ungerboeck1982channel} (TCM) which delivers excellent decoding performance in short-blocklength regime. 
TCM consists of a TBCC and a mapping function that preserves the magnitude distribution and generates channel signals with capacity-achieving distribution.
These two properties are proved in this paper. Simulation results show that, for the proposed CRC-TCM-PAS system with 87 input bits and 65-67 8-amplitude-modulation (8-AM) coded output symbols, the decoding performance under AWGN channel achieves the RCU bound with proper CRC and convolutional code. 

The remainder of this paper is organized as follows: Section \ref{sec: DM} reviews the DM and shows that SMDM outperforms CCDM in short-blocklength regime. Section \ref{sec: crc} proves that with non-uniform input data, CRC bits have uniform distribution. Section \ref{sec: TCM} introduces the encoding of TBCC and labeling of mapping function. This section proves that output symbols of proposed CRC-TCM-PAS system have capacity-achieving distribution. Simulation results are shown in Section \ref{sec: simu} and Section \ref{sec: conc} concludes this paper.

\section{Distribution Matching}\label{sec: DM}
Fig. \ref{fig: transmitter} illustrates the diagram of CRC-TCM-PAS transmitter. The transmitter consists of three key modules: distribution matcher, CRC bit generator and TCM. This section describes distribution matcher in detail and shows that for short-blocklength regime, SMDM is a better choice than CCDM.

Let $S$ be a random variable that obeys $\text{Bernoulli}(\frac{1}{2})$, and $A$ be a random variable with alphabet $\mathcal{A}=\{0,1,...,|\mathcal{A}|-1\}$. Denote the random sequence of $S$ with length $k$ by $S^k$, and the random sequence of $A$ with length $N$ by $A^N$. Specifically, $S^{k}=[S_1,\dots,S_k]$ and $A^{N}=[A_1,\dots,A_N]$. A \textit{fixed-to-fixed}  distribution matcher is an invertible function $f_{DM}$ that maps a length-$k$ source binary sequence $S^k$ to a length-$N$ sequence $A^N$:
\begin{align}
    f_{DM}: \{0,1\}^k\rightarrow \mathcal{A}^N.  
\end{align}
Denote the range of $f_{DM}$ by the codebook $\mathcal{C}_{DM}$ and note that $ \mathcal{C}_{DM}\subseteq  \mathcal{A}^N$.
The goal of a distribution matcher is that, in average, the distribution of output symbols of a DM, $P(\bar{A})$, is as close as possible to the desired distribution $P(\hat{A})$. One metric to measure the performance of a distribution matcher is normalized KL divergence \cite{CCDM_schulte}, which is defined as: 
\begin{align}\label{equ: normalized_kl}
\begin{split}
        &\frac{\mathbb{D}_{KL}\left(P(A^N)||P(\Tilde{A}^N)\right)}{N}\\
    =&\frac{1}{N2^k}\sum_{a^N\in \mathcal{C}_{DM}}\log\frac{1}{P_{\hat{A}^N}(a^N)}-\frac{k}{N}
\end{split}.
\end{align}
\eqref{equ: normalized_kl} indicates that the codebook of the optimal distribution matcher should consist of  the first $2^k$ length-$N$ sequences obtained by sorting all possible codewords in an ascending order with respect to  $\log\frac{1}{P_{\hat{A}^N}(a^N)}$, and this optimal DM is called shell-mapping DM (SMDM)\cite{SMDM_amjad}. The other well-known DM is constant composition DM (CCDM), whose codebook contains the sequences that have the same portion of $a\in\mathcal{A}$. CCDM has been proven to be asymptotically optimal. The following example shows that for short blocklength, SMDM delivers smaller normalized KL divergence than CCDM.
\begin{example}
Given the desired blocklength $N$=64 and distribution $P(\hat{A})=\{0.587,0.312,0.014,0.085\}$, the SMDM codebook has cardinality $|\mathcal{C}_{SMDM}|=2^{87}$ with normalized KL divergence 0.0376 bits, whereas CCDM codebook has cardinality $|\mathcal{C}_{SMDM}|=2^{79}$ with normalized KL divergence 0.1335 bits.
\end{example}

Example 1 shows that SMDM can provide more information with smaller divergence for short $N$. Considering our short blocklength target, this paper uses SMDM as DM module. 

\section{Cyclic Redundancy Check Encoding}\label{sec: crc}
The binary converter maps a symbol sequence $a^N\in\mathcal{C_{DM}}$ to a binary sequence. Let $|\mathcal{A}|$ be some power of 2, i.e., $|\mathcal{A}|=2^{\alpha}$. For any $a\in\mathcal{A}$,  let $\mathbf{b}(a)=[w_{\alpha}...w_{2}w_{1}]\in\mathbb{F}_2^{\alpha}$. The non-uniformity of $A$ results in different distribution for each bit $w_i$. Given $P(A)$, the PMF of $i^{th}$ bit in $\mathbf{b}(A)$, $P(W_i)$ is calculated by:
\begin{align}\label{equ: distribution}
    P_{W_i}(w)=\sum_{a=1}^{|\mathcal{A}|-1}P_A(a)\mathbb{I}\left(\mathbf{b}^{i}(a)=w\right),~w=0,1.
\end{align}
Denote the binary representation of ${a}^N$ by $w^{N\alpha}$, which is also represented in polynomial form by $w(x)=\sum_{i=0}^{N\alpha-1}w_ix^i\in\mathbb{F}_2[x]$, where $w_i$ is $i^{th}$ bits in $w^{NL}$ and $\mathbb{F}_2[x]$ denotes binary polynomial.
A $m$-bit CRC is specified by a degree-$m$ binary polynomial $p(x)=\sum_{i=0}^mp_{i}x^i$. 
Let $u(x)=\sum_{i=0}^{N\alpha+m-1}u_ix^i$ be the output of CRC encoder, $u(x)$ is calculated by:
\begin{align}
    u(x)=x^mv(x)+v(x)(\text{mod } p(x)).
\end{align}
The CRC code is systematic, as $u_{i+m}=w_{i}$, for $i=0,...,N\alpha-1$. 
Denote random sequence of CRC code output by $U^{N\alpha+m}$, it has $P(U_{i+m})=P(W_i)$. $i=0,...,N\alpha-1$. The following theorem shows that $P(U_i)$, $i=0,...,m-1$ have uniform distribution for the $\alpha=2$ case.

\begin{figure*}[t] 
\centering
  \includegraphics[scale = 0.6]{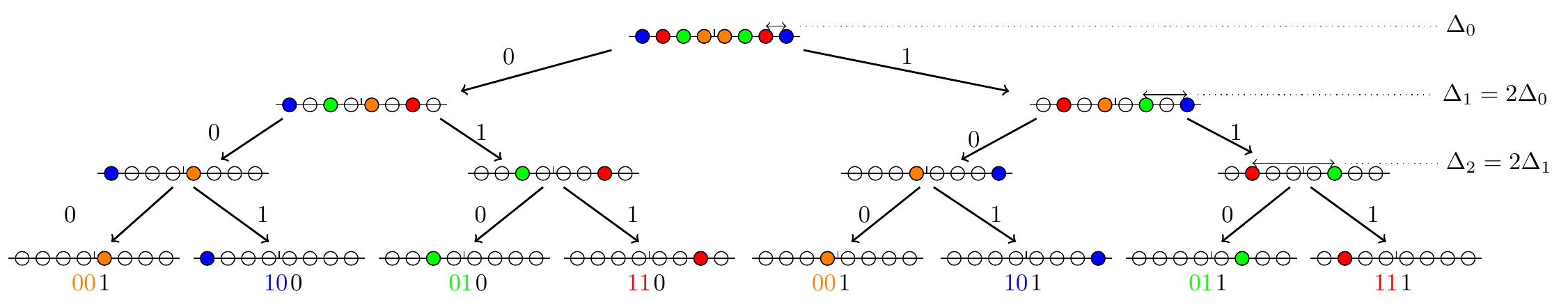}
  \caption{Partitioning of 8-AM channel signals into subsets with increasing minimum subset distance $\Delta_0<\Delta_1<\Delta2$}
  \label{fig: 8-AM-set-partition}
\end{figure*}

\begin{theorem}\label{the: uniform_bits}
For a length-$N$ random sequence $A^N$ whose elements $A_i$, $i=0,...,N-1$, are identical independent distribution (i.i.d) random variables with alphabet $|\mathcal{A}|=\{0,1,2,3 \}$ and distribution $P(A)$. Let $W^{2N}$ be the binary representation of $A^N$ and $U^{2N+m}$ be the CRC output sequence by encoding $W^{2N}$  with some degree-m CRC polynomial $p(x)$, When $N$ is large enough, i.e, when $N\rightarrow \infty$, it has:
\begin{align}
        P_{U_i}(u) &= 
     \left\{ \begin{array}{l l} 0.5,   & ~ u=0 \\  0.5, &  ~ u=1 \\ \end{array}, \right.
\end{align}
for $i=0,1,...,m-1$.
\end{theorem}
\begin{proof}
Define set $\mathcal{W}_e=\{W_{2i},i=0,...,N-1\}$ and $\mathcal{W}_o=\{W_{2i+1},i=0,...,N-1\}$. Since $|\mathcal{A}|=4$, based on \eqref{equ: distribution}, the random variables in the same set have same distribution. We specify the probability that $W_i$ is 0 as  follows:
\begin{align}
        P_{W_i}(0) &= 
     \left\{ \begin{array}{l l} p_e,   & \text{if } W_i\in \mathcal{W}_e \\  p_o, &  \text{if } W_i\in \mathcal{W}_o \\ \end{array}, \right.
\end{align}
and $P_{W_i}(1)=1-P_{W_i}(0)$. 

A CRC code is a linear block code. Denote the set of information bits constrained by $i^{th}$ parity check by $\mathcal{W}_i$, 
where $i=0,...,m-1$. Let $J_{i,e}$ be the number of the elements belonged to both $\mathcal{W}_e$ and $\mathcal{W}_i$,  and $J_{i,o}$ be the number of elements belonged to both $\mathcal{W}_o$ and $\mathcal{W}_i$ . The p.m.f of $i^{th}$ parity check bit, $P(U_i)$,  can be calculated by:
$P(U_i)=\circledast_{W_j\in\mathcal{W}_i}P(W_j)$, where $\circledast$ is the notation of circular convolution. The discrete Fourier transform (DFT) of $P(U_i)$ is calculated by 
\begin{align}
    DFT(P(U_i))&=\prod_{W_j\in\mathcal{W}_i}DFT(P(W_j))\\
    &=[1~~ (1-2p_e)^{J_{i,e}}(1-2p_o)^{J_{i,o}}].
\end{align}
By implementing inverse DFT,  $P(U_i)$ is given by
\begin{align}
     P_{U_i}(u_i) &= \frac{1}{2}+(-1)^{u^i}\frac{1}{2}(1-2p_e)^{J_{i,e}}(1-2p_o)^{J_{i,o}},
\end{align}
When $N$ is large, $J_{i,e}$ and $J_{i,o}$ are also large. Since $|1-2p_e|<1$,  $P(U_i)$, $i=0,...,m-1$,  obey uniform distribution, when $N$ is large.
\end{proof}



\begin{remark}
Theorem \ref{the: uniform_bits} uses $|\mathcal{A}|=4$ at convenience. This theorem can be generalized to any $|\mathcal{A}|=2^\alpha$ case and any linear block code, such as the observation in \cite{pas_bocherer_1} for LDPC code. Besides, if $m$ is divisible by $\alpha$, the $\frac{m}{\alpha}$ random variables corresponding to the CRC bits have uniform distribution with alphabet $\mathcal{A}$. 
\end{remark}



\section{Trellis Coded Modulation}\label{sec: TCM}
TCM \cite{ungerboeck1982channel} is a bandwidth efficient modulation technique that combines a convolutional code with modulation in one function. 
As an example shown in Fig. \ref{fig: transmitter}, a coded 8-AM modulation consists a rate-$\frac{2}{3}$ convolutional code followed by a mapping function. The convolution code takes a binary sequence of length $2(N+\frac{m}{2})$ as input and generates the binary output sequence of length $3(N+\frac{m}{2})$. Then, each 3 parallel bits are mapped to an 8-AM channel signal $x_i\in\mathcal{X}$, where $\mathcal{X}$ is the constellation set.

\subsection{Set Partitioning Mapping Rule}
In order to maximize free Euclidean distance (ED) of TCM, Ungerboeck in \cite{ungerboeck1982channel} proposed a mapping rule called "mapping by set partitioning". The mapping rule follows from successive partitioning of a channel-signal set into subsets with increasing minimum distance $\Delta_0<\Delta_1<\Delta_2\dots$ between the signals in these subsets. 

Fig. \ref{fig: 8-AM-set-partition} shows an example of set partitioning for an equidistant 8-AM constellation set. The partition result makes sure that $\Delta_0<\Delta_1<\Delta_2$. The other feature of the mapping in Fig. \ref{fig: 8-AM-set-partition} is that the first two bits indicate the magnitude of channel signal and the least significant bit (LSB) serves as sign indicator.
As a result, if the convolutional code is systematic, the information bits and check bits are mapped to magnitudes and signs, respectively.
Different with the PAS architecture in \cite{pas_bocherer_1}, which uses bit 1 to indicate $+1$ and use bit 0 to indicate $-1$, there is no deterministic relationship between the LSB and sign value in the labeling example shown in Fig. \ref{fig: 8-AM-set-partition}.

A systematic approach to search optimal convolutional codes maximizing the free ED is presented in \cite{ungerboeck1982channel}. However, the convolutional codes listed in \cite{ungerboeck1982channel} may not be optimal for CRC-TCM-PAS system, because the input bits to convolutional codes are not uniform. Simulation results in Section \ref{sec: simu} show that convolutional codes in \cite{ungerboeck1982channel} also deliver excellent performances in CRC-TCM-PAS. It will be our future work to find optimal convolutional code for CRC-TCM-PAS.

As shown in \cite{xiao2021finite}, the rate-achieving PMF of an AM signal  $X$ under AWGN channel should be symmetric:
\begin{align}\label{equ: symmetric_dist}
    P_{X}(x)=P_{X}(-x), x\in\mathcal{X}.
\end{align}
The next sub-section proves that \eqref{equ: symmetric_dist} holds when the convolutional code is tail-biting with a large $N$.

\subsection{Tail Biting Convolution Code}

A $\nu$-memory-elements convolutional code which takes ${k}_0$-bit input and generates $n_0$-bit output  in each stage is denoted by a $({n}_0,{k}_0,\nu)$ convolutional code. We call each $k_0$ input bits as a \textit{data frame}, and each $n_0$ output bits as a \textit{code frame}.  This paper is focused on $(k_0+1,k_0,\nu)$ convolutional code.  Let $\mathcal{U}=\{0,1,\dots,2^{k_0}-1\}$ be the set of input symbol and $\mathcal{L}=\{0,1,..,2^{n_0}-1\}$ be the set of output symbol. Denote the input symbol and output symbol in stage $t$ by $u_t$ and $l_t$, respectively.

A convolutional code with $N$ data frames can be described as a $N$-stages trellis. Denote the set of vertices (or states) at time instant $t$ by $\mathcal{V}_t$. For the convolutional code, the vertex sets at different time instant are the same, i.e.,
$\mathcal{\mathcal{V}}_t=\mathcal{V}=\{0,1,...,2^\nu -1 \}$. 
In stage $t$ denote the edge that starts with $v_t$, ends at $v_{t+1}$ and has a output $l_t$ by a  3-tuple $(v_t,l_t,v_{t+1})$. 
Let $E_t$ be the set of edges in stage $t$. In this paper, we consider the case where set of edges in all stages are the same, i.e., $E_t=E$.
Let the sequence ($v_0$, $l_0$, $v_1$, $l_1$, ..., $l_{v-1}$, $v_{N}$) be a valid path in $T$, i.e., $(v_t,l_t,v_{t+1})\in E$, $t=0,1,..., {N-1}$. A tail-biting trellis requires $v_0=v_{N}$.  

Denote the input vector in stage $t$ by $\mathbf{u}_t\in\mathbb{F}_2^{k_0\times1}$, and denote the state vector in time instant $t$ by $\mathbf{v}_t\in\mathbb{F}_2^{\nu\times 1}$. Based on the state-space representation of convolutional code\cite{weiss2001code,fragouli1999convolutional}, the $\mathbf{v}_{t+1}$ is a function of $\mathbf{v}_t$ and  $\mathbf{u}_t$, i.e., $\mathbf{v}_{t+1}=\mathbf{A}\mathbf{x_t}+\mathbf{B}\mathbf{u}_t,$where $\mathbf{A}\in\mathbb{F}_2^{\nu\times\nu}$ and $\mathbf{B}\in\mathbb{F}_2^{\nu\times k_0}$. One question for tail-biting convolutional code is that, given an input sequence $\{\mathbf{u}_t, t=0,\dots,N-1\}$, find the starting state $v_0$ such that the path has $v_0=v_N$ , which also means $\mathbf{v}_0=\mathbf{v}_N$. \cite{weiss2001code} pointed that the $\mathbf{v}_0$ can be determined by the following equation:
\begin{align}\label{equ: initial_state}
    \mathbf{v}_0=(\mathbf{A}^{N}+\mathbf{I}_\nu)^{-1}\mathbf{v}_N^{[zs]},
\end{align}
where $I_\nu$ is a size $\nu$ identity matrix and $\mathbf{A}^N+\mathbf{I}_{\nu}$ is an invertible matrix\cite{weiss2001code}. The term $\mathbf{v}_N^{[zs]}$ is called zero-state solution and is the final state when the starting state is $\mathbf{0}$ and input sequences are $\{\mathbf{u}_t, t=0,\dots,N-1\}$. The encoding process of tail-biting convolutional code has two steps: 
\begin{enumerate}
    \item Run encoding process first time by setting $\mathbf{x}_0=0$ and record $\mathbf{x}_N^{[zs]}$.
    \item Run encoding process second time by setting $\mathbf{x}_0$ using \eqref{equ: initial_state} and generate output data.
\end{enumerate}

Therefore, in order to study distribution of output data of TBCC, we first analyze the case where the initial state is zero state, and then analyze the case where the initial state is tail-biting state.

Based on the analysis on distribution matcher and CRC encoding, the data frames except from the ones corresponded to CRC bits have non-uniform distribution. Because the data frames are random variables, the state in time instant $t$, $V_t$ , is random variable. Inspired by the work in \cite{weiss2001code,fragouli1999convolutional},  this subsection uses state-space representation of convolution code to analyze the PMF of $V_t$. Define $\mathcal{P}(v_i)=\{(v_j,l)|v_j\in\mathcal{V},l\in\mathcal{L},(v_i,l,v_j)\in E\}$ Based on the trellis description, the PMF of state in time instant $t$, $V_t$, is calculated by:

\begin{align}\label{equ: state_equ}
     P_{V_t}(v_t)
    &=\sum_{v_{t-1}\in\mathcal{V}}P(v_{t-1})\sum_{(v_t,l)\in\mathcal{P}(v_{t-1}) }P(l_t,v_{t}|v_{t-1}).
\end{align}
Note that each edge is uniquely mapped to an input. Let $ {u}= g^{-1}\left((v,l,v')\right)\in \mathcal{U}$
if $(v_{t-1},l_t,v_t)$ is the edge corresponding to the starting state $v_t$ and input data frame $\mathbf{u}_t$.
 Hence, $P(l_t,v_t|v_{t-1})$ can be obtained by the data frame distribution $P_U\left(g^{-1}\left(\left(v_{t-1},l_t,v_t\right)\right)\right)$.

Define the matrix $\mathbf{C}_{t-1}\in\mathbb{R}^{|\mathcal{V}|\times|\mathcal{V}|}$ as follows:
\begin{align}
    \mathbf{C}_{t-1}(v_{i},v_{j})=\sum_{(v_{j},k)\in \mathcal{P}({v_i})} P(k,v_{j}|v_{i}),
\end{align}
Let $\mathbf{p_t}=\left[P_{V_t}(0)\dots P_{V_t}(2^{\nu}-1)\right]^T$, \eqref{equ: state_equ} can be rewritten by:
\begin{align}\label{equ: state_pmf}
    \mathbf{p}_t=\mathbf{C}_{t-1}\mathbf{p}_{t-1}= \left(\prod_{i=0}^{t-1} \mathbf{C}_{i}\right)\mathbf{p}_0.
\end{align}

\begin{theorem}\label{the: uinform_state}
For an $N$ data frame convolutional code with any initial state distribution $P(V_0)$. If the data frames are i.i.d random variables with p.m.f. $P(U)$ and $P_U(u)>0$ for $u\in\mathcal{U}$, then state distribution at time instant $N$, $P(V_N)$, is asymptotically uniform, i.e.
\begin{align}
    \lim_{N\rightarrow \infty}P_{V_N}(v_N)=\frac{1}{2^v},~\forall v_N\in\mathcal{V}.
\end{align}
\end{theorem}





\begin{proof}
The assumption that $N$ data frames have same distribution implies that $\mathbf{C}_t=\mathbf{C}$, for $t=0,...,N-1$. Hence, \eqref{equ: state_pmf} can be rewritten by   
\begin{align}\label{equ: state_pmf_2}
    \mathbf{p}_{N}=\mathbf{C}^{N}\mathbf{p}_0.
\end{align}
\eqref{equ: state_pmf} implies that $\mathbf{C}$ is a right stochastic matrix. Besides, $\mathbf{C}$ is also a regular matrix. The definition of regular matrix implies that there exists a path with finite steps for any $v_1,v_2\in\mathcal{V}$. Note that,  $\mathbf{C}$ contains structure of trellis $T$, i.e., for any $v_i,v_j\in\mathcal{V}$ and some  $l\in\mathcal{L}$, if $(v_i,l,v_j)\in E$, then $\mathbf{C}(v_i,v_j)\neq 0$. For the convolutional code considered in this paper, $v_i$ can always reach $v_j$ with finite stages. As a result, $\mathbf{C}$ is a regular right stochastic matrix. Based on Perron-Frobenius theorem\cite{gantmakher2000theory}, for any regular right stochastic matrix $\mathbf{C}$, it has:
\begin{enumerate}
    \item The matrix $\mathbf{C}$ has $1$  as an eigenvalue of multiplicity 1.
    \item All the other eigenvalues $\lambda_j$ have $|\lambda_j|<1$.
\end{enumerate}

Let $\mathbf{Q}\mathbf{J}\mathbf{Q^{-1}}$ be the  Jordan Canonical form   of $\mathbf{C}$. Based on Perron-Frobenius theorem, 
$\mathbf{J}=\text{diag}(1,\mathbf{J}_2,\dots,\mathbf{J}_q),$
where $\mathbf{J}_i$, $i=2,...,q$ are Jordan block matrices with some eigenvalue which maginutde is less than $1$. Let $\mathbf{Q}=[\mathbf{q}_1\dots\mathbf{q}_{2^\nu}]$ and, $\mathbf{q}_1$ is the eigenvector of $\mathbf{C}$ with eigenvalue 1. Due to the stochastic property, the normalized eigenvector corresponding to eigenvalue 1 is $\mathbf{q}^*_1=[\frac{1}{\sqrt{2^\nu}}\dots\frac{1}{\sqrt{2^\nu}}]^T$, let $\mathbf{q}_1= r\mathbf{q}_1^*$. Let $\mathbf{p}_0=\sum_{i=1}^{2^\nu}c_i\mathbf{q}_i=\mathbf{Q}\mathbf{c}$, it has $\mathbf{p}_N=\mathbf{C}^N\mathbf{p}_0=\mathbf{Q}\mathbf{J}^N\mathbf{c}$.
Note that $\mathbf{J}_i\rightarrow \mathbf{0}$ as $N\rightarrow\infty$, therefore
\begin{align}
    \lim_{N\rightarrow \infty} \mathbf{p}_N &= c_1r\mathbf{q}_1^* =\left[\frac{1}{2^\nu}\dots \frac{1}{2^\nu}\right]^T\label{equ: final_uniform}.
\end{align}
Thus, when $N$ is large enough, $P(V_N)$ converges to uniform distribution.
\end{proof}  


Similarly, define $\mathcal{Q}(l)=\{(v_i,v_j)|v_i,v_j\in\mathcal{V},(v_i,l,v_j)\in E\}$, the PMF of output in stage $t$, $P(L_t)$, is calculated by 
\begin{align}\label{equ: output_dist}
P_{L_t}(l_t)&=\sum_{v\in\mathcal{V}}P_{V_{t-1}}(v)\sum_{(v,v_t)\in \mathcal{Q}(l)}P(l_t,v_{t}|v).
\end{align}

Define the matrix $\mathbf{D}\in\mathbb{R}^{|\mathcal{L}|\times|\mathcal{V}|}$ as follows
\begin{align}
    \mathbf{D}(l,v)=\sum_{(v,v')\in\mathcal{Q}(l)}P(l_t,v_{t}|v),
\end{align}
where $l\in\mathcal{L}$ and $v\in\mathcal{V}$. Define $\mathbf{q}_{t}=[P_{L_t}(0) ... P_{L_t}(|\mathcal{L}|-1)]^T$. Then \eqref{equ: output_dist} can be rewritten as:
\begin{align}
    \mathbf{q}_t=\mathbf{D}\mathbf{p}_t.
\end{align}

\begin{theorem}\label{the: sym_dist}

If $\mathbf{p}_t=[\frac{1}{2^\nu}\frac{1}{2^\nu}\dots \frac{1}{2^\nu}]^T$, then:
\begin{align}
    P_{L_t}(l)= \frac{1}{2} P(g^{-1}(v_{t-1},l,v_t)), 
\end{align}
for any $l\in\mathcal{L}$ and any $(v_{t-1},v_t)\in\mathcal{Q}(l)$.
\end{theorem}

\begin{proof}
The matrix $\mathbf{D}$ has two important properties. The first property is that each row contains $2^{\nu-1}$ non-zero elements. This is because that the register that is most adjacent to the output is uniquely determined by the code frame, therefore it only has $2^{\nu-1}$ possible states that ends at some states with the given output. The second property is that the non-zeros elements in each row have same values.  This property comes from the fact that the considered convolution code has systematic form and therefore each output $l_t$ corresponding to the edge $(v,l_t,v')$  is uniquely mapped to the input $g^{-1}(v,l_t,v')$. Therefore, for any $l\in\mathcal{L}$,  it has:
\begin{align}
P_{L_t}(l) &= \sum_{i=1}^{2^\nu}\mathbf{D}(l,i)P_{V_t}(i),\\
&=\frac{1}{2^\nu}2^{\nu-1}P(g^{-1}(v_{t-1},l,v_t)),\\
&=\frac{1}{2}P(g^{-1}(v_{t-1},l,v_t)).
\end{align}
\end{proof}

\begin{remark}
Let $\mathbf{b}(l)$ be the binary representation of $l$ and $l(0)$ be the LSB of $\mathbf{b}(l)$.  Because the convolutional code is systematic, $l(0)$ is the check bits corresponded $l$. Theorem \ref{the: sym_dist} implies that,  for $l$, $l'$, whose binary representations only differ in parity check bit, it has
\begin{align}\label{equ: equal_prob}
P_{L_N}(l)=P_{L_N}(l').    
\end{align}

\end{remark}

Finally, for the CRC-TCM-PAS system, in order to generate tail-biting path, the $N+\frac{m}{\alpha}$ input data frames are encoded with initial state zero. Note that the first $N$ data frames have same distribution, and $N$ is large enough such $V_{N}$ has uniform distribution. The last $\frac{m}{\alpha}$ symbols have uniform distribution. With \eqref{equ: state_pmf}, it is easy to show that  $V_{N+\frac{m}{l}}$ has uniform distribution.

As indicated in \eqref{equ: initial_state}, the TBCC initial state is a linear transformation of $V_{N+\frac{m}{l}}$, thus the initial state of TBCC in CRC-TCM-PAS has uniform distribution. This also implies that the states in  all $N+1$ time instants in TBCC have uniform distribution. Because the information bits determine magnitude and check bit determines signs of the channel signal, based on \eqref{equ: equal_prob}, we have that for the output channel signals of CRC-TCM-PAS system, $\{X_i,i=0,...,N+\frac{m}{l}-1\}$,
    $P_{X_i}(x)=P_{X_i}(-x)$, $x\in \mathcal{X}$. 
With proper design for distribution matcher, the first $N$ symbols have capacity-achieving distribution and last $\frac{m}{\alpha}$ symbols have uniform distribution.

\subsection{List Decoding}
Under the AWGN channel, Viterbi algorithm finds the codewords that has minimum Euclidean distance to the channel observation. In TBCC-CRC-PAS system, the prior of channel signaling must be taken into consideration. Let $\mathbf{y}\in\mathbb{R}^N$, the Viterbi algorithm finds $\mathbf{x}^*\in\mathcal{X}^N$ such that
\begin{align}
    \mathbf{x}^*= \arg\min_{\mathbf{x}\in \mathcal{X}^N}\sum_{i=1}^N\left[(x_i-y_i)^2+2\sigma^2\log\frac{1}{P_X(x_i)}\right].
\end{align}
The serial list Viterbi decoding (S-LVD) \cite{seshadri1994list} sequentially  finds the first $H$ most likely codewords. With CRC concatenated, S-LVD works as follows: S-LVD first finds the most likely codeword and passes it through CRC check. If the codeword passes the CRC check then S-LVD claims a success and stops. Otherwise, S-LVD finds the second most likely codewords and passes it to CRC check. The decoding process is repeated until $H^{th}$ most likely codeword are searched. 


\begin{figure}[t] 
\centering
  \subfloat[{}\label{fig: m=3}]{%
       \includegraphics[width=20pc]{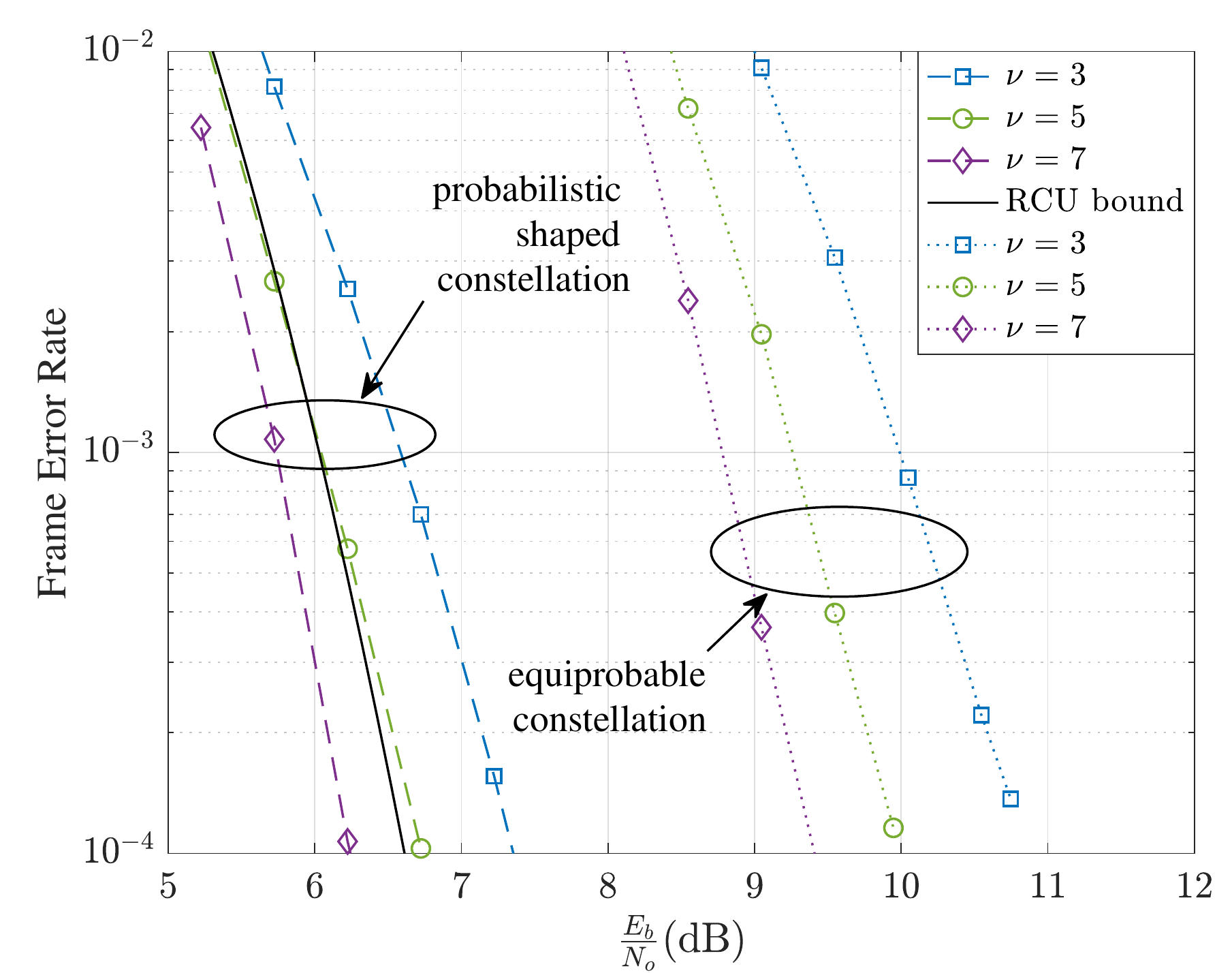}}
\hfill
  \subfloat[\label{fig: m=7}]{%
        \includegraphics[width=20pc]{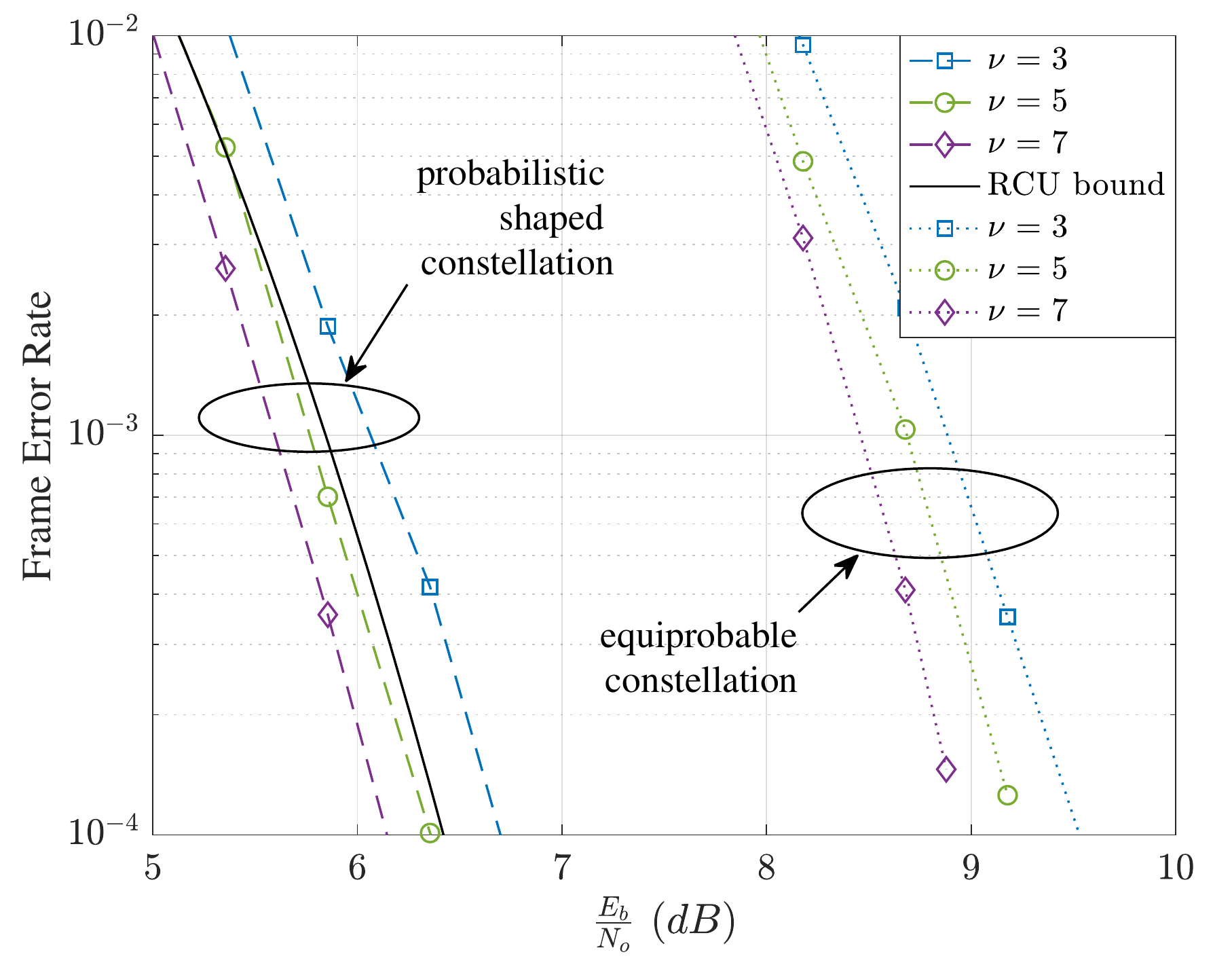}}

 \caption{The FER curves of CRC-TCM-PAS transmission system with 8-AM coded modulation and : a) degree 2 CRC and b) degree 6 CRC. The black cuuves are RCU bound for CTC-TCM-PAS with corresponding CRC length. The FER performances of the system without DM module, i.e., CRC-TCM, are provided for comparison. }
\label{fig: FER}

\end{figure}

\section{Simulation Result}\label{sec: simu}
In this section, we exam the performance of the proposed CRC-TCM-PAS system under AWGN channel. In this paper, we consider the channel signal as 8-AM symbols with equidistance. The constellation set $\mathcal{X}$ and corresponding PMF are optimized using dynamic-assignment Blahut-Arimoto
algorithm\cite{xiao2021finite}. 

For the TBCC-CRC-PAS transmitter, SMDM takes $k=87$ bits as input and output $N=64$ symbols with average PMF $P({\bar{A}})=[0.5742,0.3188,0.01642,0.09048]$. The convolutional codes with different memory elements $\nu$ are taken from \cite{ungerboeck1982channel}. Finally, CRC polynomials are searched in a brute force manner, all the CRCs are simulated and the one delivers best performance is chosen. Note that there are efficient CRC selection algorithms for tail-biting convolution code\cite{9174422}.  On the decoder side, the list decoder has list size $H=30$.
  \begin{figure}[t]
	\centering
	 \includegraphics[width=20pc]{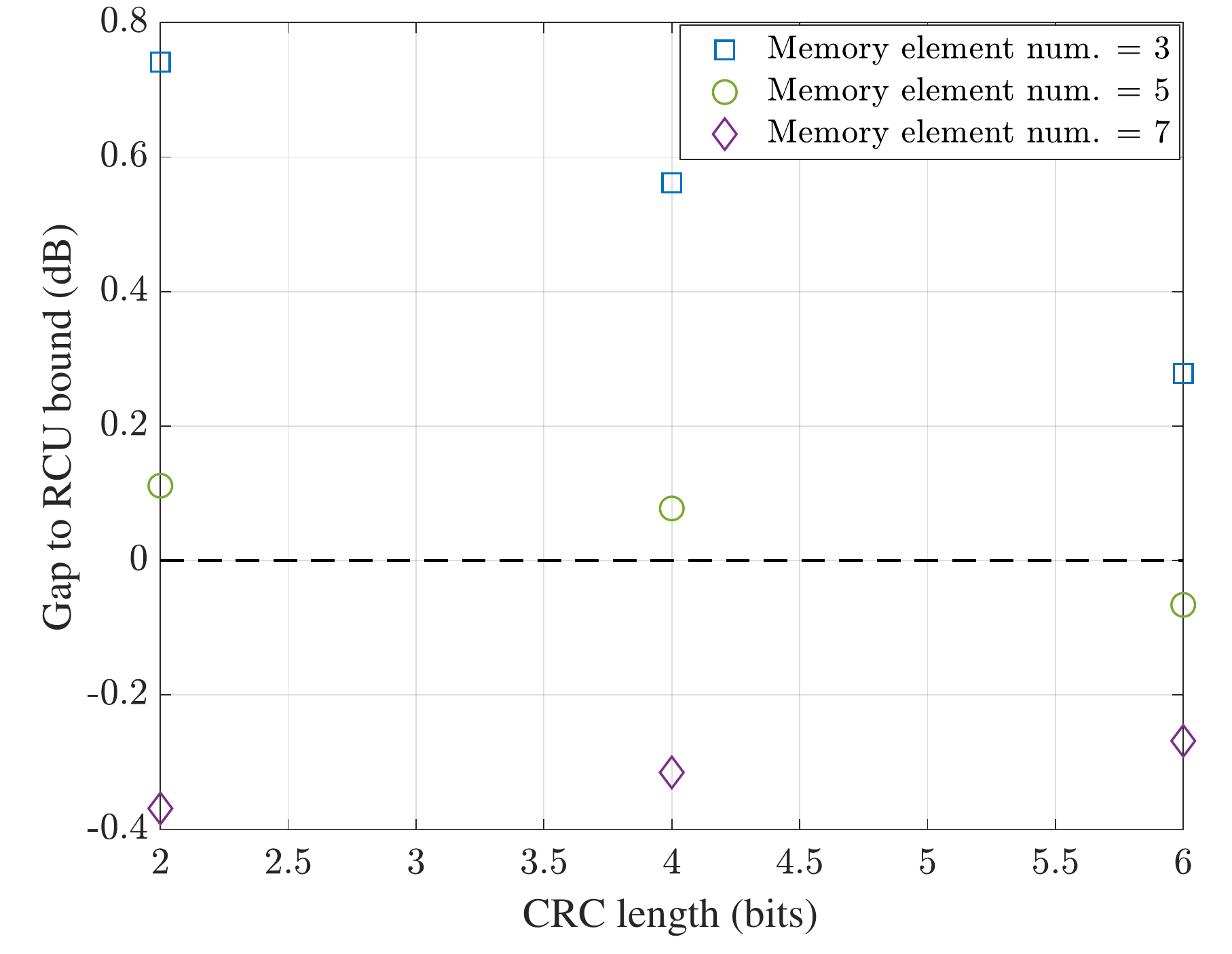}
	 \caption{The gaps of TBCC-CRC-PAS system with various CRC length and number of elements $\nu$ to the RCU bound at the FER of $10^{-4}$. The negative values indicate the dB values which TBCC-CRC-PAS systems outperforms RCU bound at the FER of $10^{-4}$.}
    \label{fig: RCU_gap} 
\end{figure}
Fig. \ref{fig: m=3} shows the frame error rate (FER) of CRC-TCM-PAS system with 2 CRC bits and various number of elements, $\nu$, for the convolutional code. As a comparison, the RCU bound is provided in Fig. \ref{fig: m=3}. As indicated in Sec. \ref{sec: simu}, the output symbols of the CRC-TCM-PAS don't have same distribution. The first $N$ symbols obey the DM output distribution and last $\frac{m}{2}$ symbols have uniform distribution. In order to calculate RCU bound, we assume all output symbols have PMF $P(\bar{A})$. Simulation result shows that the decoding performance gets improved with the increase of $\nu$. With $\nu=7$, the FER performance is better than the RCU bound. Fig. \ref{fig: m=7} shows the FER curves with 6 CRC bits and various memory elements, hence the output blocklength is 67, which has a lower transmission rate compared with 2-bit CRC system. Simulation results shows that when the CRC-TCM-PAS system implements 6-bit CRC, the FER can outperform the RCU bound only with $\nu=5$. 

Fig. \ref{fig: FER} also provides the decoding performances of the system with equiprobable constellation for comparison. The CRC-TCM takes $128$ binary bits as input and generates $64+\frac{m}{2}$ 8-AM output symbols. Simulation results show that the system with DM has a better decoding performance.

Fig. \ref{fig: RCU_gap} shows the gaps of CRC-TCM-PAS system with various CRC length and number of elements $\nu$ to the RCU bound at the FER of $10^{-4}$. The negative values indicate the dB values which CRC-TCM-PAS systems outperform RCU bound at the FER of $10^{-4}$. Simulation result shows that increasing $\nu$ improves the gap greatly. For $\nu=7$, the CRC-TCM-PAS system with all investigated CRC length outperforms RCU bound. Fig. \ref{fig: RCU_gap} also shows that, increasing the number of elements for shorter CRC length has a larger benefit on decoding performance than for longer CRC length.

\section{Conclusion}\label{sec: conc}
This paper proposes CRC-TCM-PAS which applies probabilistic amplitude shaping  to a cyclic redundancy check  aided trellis coded modulation to achieve the short-blocklength RCU bound. 
This paper proves that with non-uniform input data, CRC bits have uniform distribution. 
This paper also proves that output symbols of proposed CRC-TCM-PS system have capacity-achieving distribution. Simulation results show that, for the proposed PS-CRC-TCM system with 87 input bits and 65-67 8-AM coded output symbols, the decoding performance under AWGN can achieve RCU bound, when proper CRC and convolutional code are selected.

\bibliographystyle{IEEEtran}
\bibliography{Ref}

\begin{thebibliography}{10}
\providecommand{\url}[1]{#1}
\csname url@samestyle\endcsname
\providecommand{\newblock}{\relax}
\providecommand{\bibinfo}[2]{#2}
\providecommand{\BIBentrySTDinterwordspacing}{\spaceskip=0pt\relax}
\providecommand{\BIBentryALTinterwordstretchfactor}{4}
\providecommand{\BIBentryALTinterwordspacing}{\spaceskip=\fontdimen2\font plus
\BIBentryALTinterwordstretchfactor\fontdimen3\font minus
  \fontdimen4\font\relax}
\providecommand{\BIBforeignlanguage}[2]{{%
\expandafter\ifx\csname l@#1\endcsname\relax
\typeout{** WARNING: IEEEtran.bst: No hyphenation pattern has been}%
\typeout{** loaded for the language `#1'. Using the pattern for}%
\typeout{** the default language instead.}%
\else
\language=\csname l@#1\endcsname
\fi
#2}}
\providecommand{\BIBdecl}{\relax}
\BIBdecl

\bibitem{pas_bocherer_1}
G.~B{\"o}cherer, F.~Steiner, and P.~Schulte, ``Bandwidth efficient and
  rate-matched low-density parity-check coded modulation,'' \emph{IEEE Trans.
  on comm.}, vol.~63, no.~12, pp. 4651--4665, 2015.

\bibitem{pas_bocherer_2}
G.~B{\"o}cherer, P.~Schulte, and F.~Steiner, ``Probabilistic shaping and
  forward error correction for fiber-optic communication systems,''
  \emph{Journal of Lightwave Technology}, vol.~37, no.~2, pp. 230--244, 2019.

\bibitem{pas_achievablerate_bocherer}
G.~B{\"o}cherer, ``Achievable rates for probabilistic shaping,'' \emph{arXiv
  preprint arXiv:1707.01134}, 2017.

\bibitem{polar_ook_thomas}
T.~Wiegart, F.~Steiner, P.~Schulte, and P.~Yuan, ``Shaped on–off keying using
  polar codes,'' \emph{IEEE Communications Letters}, vol.~23, no.~11, pp.
  1922--1926, 2019.

\bibitem{SMDM_amjad}
R.~A. Amjad and I.~G. B{\"o}cherer, ``Algorithms for simulation of discrete
  memoryless sources,'' Ph.D. dissertation, Master’s thesis, Technische
  Universit{\"a}t M{\"u}nchen, 2013.

\bibitem{ccdm_smdm_comp_schulte}
P.~Schulte and F.~Steiner, ``Divergence-optimal fixed-to-fixed length
  distribution matching with shell mapping,'' \emph{IEEE Wireless
  Communications Letters}, vol.~8, no.~2, pp. 620--623, 2019.

\bibitem{CCDM_schulte}
P.~Schulte and G.~B{\"o}cherer, ``Constant composition distribution matching,''
  \emph{IEEE Trans. on Info. Theory}, vol.~62, no.~1, pp. 430--434, 2015.

\bibitem{DMReview}
Y.~C. G{\"u}ltekin, T.~Fehenberger, A.~Alvarado, and F.~M. Willems,
  ``Probabilistic shaping for finite blocklengths: Distribution matching and
  sphere shaping,'' \emph{Entropy}, vol.~22, no.~5, p. 581, 2020.

\bibitem{LDPC_capacity}
T.~Richardson and R.~Urbanke, ``The capacity of low-density parity-check codes
  under message-passing decoding,'' \emph{IEEE Transactions on Information
  Theory}, vol.~47, no.~2, pp. 599--618, 2001.

\bibitem{Liang_tbcccrc}
E.~Liang, H.~Yang, D.~Divsalar, and R.~D. Wesel, ``List-decoded tail-biting
  convolutional codes with distance-spectrum optimal {CRC}s for 5g,'' in
  \emph{2019 IEEE Glob. Comm. Conf. (GLOBECOM)}, 2019, pp. 1--6.

\bibitem{yang2018serial}
H.~Yang, S.~V. Ranganathan, and R.~D. Wesel, ``Serial list viterbi decoding
  with {CRC}: Managing errors, erasures, and complexity,'' in \emph{2018 IEEE
  Glob. Comm. Conf. (GLOBECOM)}.\hskip 1em plus 0.5em minus 0.4em\relax IEEE,
  2018, pp. 1--6.

\bibitem{yang2019list}
H.~Yang, E.~Liang, H.~Yao, A.~Vardy, D.~Divsalar, and R.~D. Wesel, ``A
  list-decoding approach to low-complexity soft maximum-likelihood decoding of
  cyclic codes,'' in \emph{2019 IEEE Global Communications Conference
  (GLOBECOM)}.\hskip 1em plus 0.5em minus 0.4em\relax IEEE, 2019, pp. 1--6.

\bibitem{ungerboeck1982channel}
G.~Ungerboeck, ``Channel coding with multilevel/phase signals,'' \emph{IEEE
  trans. on Info. Theory}, vol.~28, no.~1, pp. 55--67, 1982.

\bibitem{xiao2021finite}
D.~Xiao, L.~Wang, D.~Song, and R.~D. Wesel, ``Finite-support
  capacity-approaching distributions for awgn channels,'' in \emph{2020 IEEE
  Information Theory Workshop (ITW)}.\hskip 1em plus 0.5em minus 0.4em\relax
  IEEE, 2021, pp. 1--5.

\bibitem{weiss2001code}
C.~Wei{\ss}, C.~Bettstetter, and S.~Riedel, ``Code construction and decoding of
  parallel concatenated tail-biting codes,'' \emph{IEEE Trans. on Info.
  Theory}, vol.~47, no.~1, pp. 366--386, 2001.

\bibitem{fragouli1999convolutional}
C.~Fragouli and R.~D. Wesel, ``Convolutional codes and matrix control theory,''
  in \emph{Proceedings of the 7th International Conference on Advances in
  Communications and Control, Athens, Greece}.\hskip 1em plus 0.5em minus
  0.4em\relax Citeseer, 1999.

\bibitem{gantmakher2000theory}
F.~R. Gantmakher, \emph{The Theory of Matrices, Volume 2}.\hskip 1em plus 0.5em
  minus 0.4em\relax American Mathematical Soc., 2000, vol. 133.

\bibitem{seshadri1994list}
N.~Seshadri and C.~Sundberg, ``List viterbi decoding algorithms with
  applications,'' \emph{IEEE trans. on comm.}, vol.~42, no. 234, pp. 313--323,
  1994.

\bibitem{9174422}
H.~Yang, L.~Wang, V.~Lau, and R.~D. Wesel, ``An efficient algorithm for
  designing optimal {CRC}s for tail-biting convolutional codes,'' in \emph{2020
  IEEE Inter. Symp. on Info. Theory (ISIT)}, 2020, pp. 292--297.

\end{thebibliography}
\end{document}